\RequirePackage[2020-02-02]{latexrelease}
\documentclass[%
reprint,
aps,
]{revtex4-1}

\usepackage{amsmath} 
\usepackage{amssymb}
\usepackage{dcolumn}% Align table columns on decimal point
\usepackage{graphicx}
\usepackage{bm}
\usepackage{dsfont}
\usepackage{threeparttable}

\newcommand{\rset}{\stackrel{\ {\sf R}}{\leftarrow}}

\newcommand{\set}{:=}
\newcommand{\ra}{\rightarrow}

\newcommand{\N}{\mathbb{N}}

\newcommand{\R}{\mathbb{R}}
\newcommand{\scl}{{ \overset{\delta}{\approx} }}
\newcommand{\OC}{{\sf OC}}
\newcommand{\SSS}{{\cal S}}

\newtheorem{claim}{Claim}

\usepackage{url}

\begin{document} 
 
\title{On the Arrow of Time and Organized Complexity in the Universe}

\author{Tatsuaki Okamoto}

\affiliation{{NTT} \\
{3-9-11 Midori-cho, Musashino-shi, Tokyo, 180-8585 Japan}\\
{\rm \url{tatsuaki.okamoto@gmail.com}}
}

\vspace{5pt}

\date{February 26, 2026}

\begin{abstract}
There is a widespread assumption that the universe in general,
and the Earth's biosphere in particular, is becoming more complex over time.
This paper formulates this assumption as a macroscopic law,
{\it the law of increasing complexity}, for a system
over a finite time span.
It hypothesizes that this macroscopic law
emerges in certain non-equilibrium systems with abundant free energy flows, 
such as the observable universe and the Earth's biosphere.
We distinguish between two types of complexity: 
disorganized and organized. 
The complexity associated with this assumption is {\it organized complexity}.
To formulate this law, 
we employ a quantitative definition of 
organized complexity as applied to 
probability distributions.
We represent any object of complexity 
as the source of its observed value, 
which is expressed as a probability distribution; 
this enables a unified treatment of diverse objects.
This formulation necessitates the use of
observation systems to represent these objects.
We introduce an order relation between these observation systems to demonstrate that the complexity of an object possesses a generic property,
one that does not depend on any specific observation system. 

This paper develops a novel methodology for this
macroscopic law, which formulates {\it the arrow of time 
in terms of increasing organized complexity 
for certain non-equilibrium systems}.
This contrasts with the second law of thermodynamics, which
formulates the arrow of time in terms of increasing disorganized complexity 
(entropy) for isolated systems. 
The formulation of this law provides 
a theoretical foundation for the phenomena of increasing complexity 
in the universe
through its potential reduction to fundamental physical laws.

We apply this formulation to the {\it fine-tuning} problem:
the puzzling observation  
that the fundamental physical constants appear to be fine-tuned for life on Earth. 
Our new explanation of the fine-tuning problem 
posits that these constants are fine-tuned for 
{\it the emergence of the law of increasing complexity}. 
This explanation is more plausible 
than those based on the notion of fine-tuning for life 
or the anthropic principle.

\end{abstract}

\newtheorem{theorem}{Theorem} 
\newtheorem{definition}{Definition}
\newtheorem{remark}{Remark}

\maketitle

\section{Introduction}

\subsection{Emergence}
\label{sec:emergence}

We often observe macroscopic phenomena in systems %the 
composed of numerous %small 
elements that do not occur in systems with only a few
\cite{Anderson72,Laughlin05}.
For example,  thermodynamic phenomena %regarding thermodynamics 
and the phases of matter %such as the %familiar 
(e.g., liquid, vapor, and solid) 
are macroscopic phenomena that appear in large assemblies of elements. 
We call such macroscopic phenomena %the 
{\it emergence}. %in this paper.

A dissipative structure, introduced by Ilya Prigogine \cite{NicPri77}, 
is characterized by the emergence of order or self-organization
in a thermodynamic non-equilibrium open system, 
an assembly of numerous elements. %atoms and molecules
Such structures do not appear in systems with %microscopic systems. 
only a few elements. %several atoms and molecules (in a microscopic world)
They are typical examples of emergence in non-equilibrium open systems, 
where energy and matter enter or leave.
The phenomena of life, which are hierarchically composed of numerous 
interacting elements, also represent emergence 
associated with dissipative structures.

Although macroscopic phenomena emerge from fundamental physical laws, it is generally difficult to strictly reduce them to these laws
\cite{Anderson72,Laughlin05}. %Moreover,  
Instead, %macroscopic phenomena 
they are often effectively %succinctly 
explained by macroscopic theories and laws rather than fundamental laws. For instance, macroscopic heat phenomena emerge from %as 
the collective behavior of %a vast number of 
molecules obeying quantum mechanics. 
However, these macroscopic phenomena are effectively explained by the macroscopic theory of thermodynamics rather than by the underlying laws of quantum mechanics.
Therefore, fundamental laws alone %physical laws %in the microscopic world 
are insufficient to explain %various %macroscopic 
phenomena in the universe; %but
we also require emergent macroscopic laws, 
such as the second law of thermodynamics. 
 
\subsection{Arrow of Time}
\label{sec:arrow}
 
Over the 13.8 billion years since its inception, the cosmos has undergone a continuous escalation of structural complexity, originating from a nearly featureless state. 
In fact, we observe a profound flow starting from the state immediately after the Big Bang, 
when not even elementary particles existed,
through the formation of elementary particles, light atoms and protostars,
to the evolution %the emergence 
of heavy atoms, diverse molecules, planets,
stars, galaxies, and (super)clusters of galaxies.  
The genesis of Earth 4.6 billion years ago and the biogenesis 4 billion years ago initiated the emergence of various life forms, 
the development of the global biosphere, 
the appearance of the human race more than 200,000 years ago, 
and the evolution of the noosphere (civilization and society).  

Given such temporal transitions in the universe, there is a widespread assumption that

\vspace{5pt}

{\it the universe in general,
and the biosphere on Earth in particular, is becoming more complex over time.} \cite{LinDavRus13}

\vspace{5pt}
 
As described above in Section \ref{sec:emergence}, 
it is natural to suggest that 
a macroscopic law may underlie these emergent phenomena.
%transitions in the universe,and 
Undoubtedly, many have already considered this idea \cite{LinDavRus13};
however, it has yet to be formalized as a rigorous, bona fide law.

This leads to a fundamental question:

\vspace{5pt}
{\it Can we formulate this assumption
as an emergent macroscopic law?}

\subsection{Contribution}

This paper addresses the aforementioned question 
as a macroscopic law, {\it the law of increasing complexity},
for a system over a finite time span.
We hypothesize that this macroscopic law
emerges in certain non-equilibrium %open 
systems with abundant free energy flows, 
such as the observable universe and the Earth's biosphere.
There are two types of complexity: disorganized and
organized \cite{Weaver48,Okamoto22}. 
The complexity associated with this assumption is {\it organized complexity}.
To define the degree of organized complexity for our formulation, 
we use a quantitative definition of the complexity of organized matters,
termed {\it organized complexity} (OC) \cite{Okamoto22} (Appendix A), %\ref{sec:oc}),
which is defined for probability distributions.
We represent any object of complexity 
as the source of its observed value, treated as a probability distribution;
this allows us to handle various objects in a unified manner.
To represent vast systems (objects), 
such as the observable universe or the Earth's biosphere, %on Earth 
as the source of their observed values,
we require observation systems consisting of a large quantity of diverse  
equipment and devices.
We assume that the configuration, encompassing %comprising %such as 
the locations and arrangements of this equipment, %and devices, 
is optimized to yield the highest possible observed complexity 
for a given object.
We then introduce an order relation ($\geq$) between observation systems
to demonstrate that the complexity of an object possesses a generic property:
%(e.g., increasing complexity over time), 
one that does not depend on any specific observation system. 

Broadly speaking, this macroscopic law 
for %an open 
a system $\SSS$ (an assembly of many elements)
states that %there exist open systems in the universe, 
$\SSS$ is associated with a time interval 
$I_\SSS \set [T_0, T_1]$ ($T_0, T_1 \in \R$, $0< T_0 < T_1$) and
an observation system $\cal{O}_{\SSS}$ %with many devices 
%$D_\SSS \set (d_1, \ldots, d_m)$ %, and a precision level $\delta_\SSS$,
such that 
%\noindent
for any observation system ${\cal O} \geq {\cal O}_{\SSS}$, %any precision level $\delta \leq \delta_\SSS$, 
and 
any time $T \in I_\SSS$, 
there exists $s > 0$ %\in \R^+$,
$$
{\sf OC}^*_{\cal{O}}(\SSS_{T+s})
>  {\sf OC}^*_{\cal{O}}(\SSS_T).
$$
Here, $I_\SSS$ and $\cal{O}_{\SSS}$ are proper for each $\SSS$, 
$\SSS_t$ is the state of %the open 
system $\SSS$ at time $t \in I_\SSS$, 
and ${\sf OC}^*_{\cal{O}}(A)$ 
is the OC (organized complexity) of
the source of the output of the observation system ${\cal O}$
%given observing 
for object $A$ with an optimized configuration. 
Note that this law does not imply that complexity 
increases monotonically; rather, 
it suggests that while complexity may temporarily decrease, 
it will inevitably increase at some later time.

The formulation of this law provides %paves the way for
a theoretical foundation for the increasing complexity phenomena 
%of increasing complexity  
in the universe
through its potential reduction to fundamental physical laws.
%As an instance, 
We apply this formulation 
to the {\it fine-tuning problem}: %of the fundamental physical constants: 
the puzzling observation that the fundamental physical constants appear to be fine-tuned for life on Earth. 
Our new explanation 
posits that 
{\it these constants 
are fine-tuned for the emergence of the %(macroscopic) 
law of increasing complexity.} 
While the creation of life on Earth is a significant phenomenon, the emergence of this macroscopic law represents a more fundamental and profound property of the universe. Furthermore, our explanation is grounded in an objectively and mathematically formulated law, whereas "life" and the "anthropic principle" are often regarded as subjective or vague notions. Thus, our explanation is scientifically more plausible than those based on the notion of fine-tuning  for life or the anthropic principle. An (approximate) reduction of this macroscopic law to fundamental laws would allow for a concrete assessment of the conditions required for this law to emerge in the universe.

\subsection{Notations}
\label{sec:notation}

$X \set Y$ denotes that 
$X$ is defined by $Y$.
The sets of natural, and real numbers
are denoted by $\N$, and $\R$, 
respectively. 
The set of $n$-bit strings is denoted by $\{0,1\}^n$ ($n \in \N$),
$\{0,1\}^* \set \cup_{n \in \N} \{0,1\}^n$.
%and the null string (0-bit string) is denoted by $\lambda$. 
When $x \in \{0,1\}^*$, $|x|$ denotes the bit length of $x$. 
When $a, b \in \R$, 
$[a, b]$ denotes set $\{x \mid x \in \R, \   
a\leq x \leq b \} \subset \R$. 
A probability distribution over $\{0,1\}^n$ is 
$\{ (a, p_a) \mid a \in \{0,1\}^n, 
p_a \in [0,1], 
%p_a \in \R, 0\leq p_a \leq 1, 
\sum_{a\in \{0,1\}^n}p_a=1 \}$. 
When $X$ and $Y$ are two distributions,  
the statistical distance of $X$ and $Y$,
${\sf SD}(X,Y)$, is defined by
$\frac{1}{2}\cdot \sum_{\alpha \in \{0,1\}^*} | \Pr[\alpha \rset X]
- \Pr[\alpha \rset Y] |$,
and
$X \overset{\delta}{\approx} Y$ 
denotes that ${\sf SD}(X,Y)$ 
is bounded by $\delta$.
When $A$ is a probability distribution,
$a \rset A$ denotes that 
element $a \in \{0,1\}^n$ is randomly selected from $A$

\section{Law of Increasing Complexity and Hypothesis} 

\subsection{Issues}
\label{sec:issues}

This section addresses %answers 
the question posed in Section %I.A. 
\ref{sec:arrow}. 
To do so, we must resolve the following issues:
\begin{itemize}
\item
Objects of complexity manifest in various forms, such as  matter, planets, stars, galaxies, % clusters of galaxies, 
life, the biosphere, %humans, 
and human societies.
Can we represent %define the degree of complexity of 
such a variety of objects in a unified manner? 

\item
What is complexity?
There are two types: disorganized and organized complexity
 \cite{Weaver48,Okamoto22}.
Which is most suitable for our purpose?
%The universe has 

How should the degree of complexity of an object be defined?
Furthermore, when an object is a massive system comprising %consisting of 
a vast number of elements %al objects 
such as the observable universe 
and the Earth's biosphere,
how do we define its complexity?

\item
Assuming that the universe's complexity increases,
does this increase occur at the ``top of the pyramid'' (i.e., the most complex level)?
We divide the question of whether complexity increases into two separate inquiries: 
(1) Is the average complexity of the universe increasing? %Moreover,
(2) Is the complexity of the universe's most complex objects or systems
%(e.g., life and humans on Earth) 
increasing?
\cite{LinDavRus13}
\item
How does an increase in organized complexity coexist 
with the second law of thermodynamics, the law of increasing entropy (disorganized complexity or randomness)?
\item
Does the complexity of a system in the universe increase monotonically?
Does the complexity of the (observable) universe 
continue to increase indefinitely?
 
\end{itemize}

\subsection{Complexity Definition for Any Objects}
\label{sec:complexity-definition-any-objects}

%\paragraph{Objects}
\subsubsection{Objects of Complexity and the Source of Observation}
\label{sec:object}

The objects of complexity encompass everything in our environment, including 
%around us, such as %elementary particles, atoms, molecules, 
matter, planets, stars, galaxies, %clusters of galaxies,  
life, the biosphere, humans, and human societies.
Our recognition of the existence of any entity relies solely on observation.
For instance, we perceive distant or microscopic objects through observations made by devices such as telescopes, microscopes, specialized apparatuses, and electronic instruments.
While we can directly interact with objects through touch, our brains ultimately perceive them %they are also recognized by our brains 
as electrical nerve signals transmitted from our sensory organs via the nervous system. 
Therefore, all objects of complexity are perceived  %recognized 
as the outcome %result 
of observations made by various  instruments, 
including human sensory organs. %the human sensors of the five senses.
In other words, we can represent any object %of complexity %treated 
as the result of its observation. 

In quantum mechanics, when observing microscopic phenomena, 
the observed value is randomly selected according to a specific %particular 
probability distribution associated with the quantum state.  
Similarly, the reception and measurement of radio signals almost always involve noise.
Chaos theory posits %states 
that fluctuations in the initial conditions can result in %lead to 
diverse phenomena that exhibit characteristics 
similar to those of random systems. 
Even such quasi-probabilistic systems can be effectively modeled as 
probabilistic systems if the initial conditions are subject to noise-induced fluctuations.
In general, an observed value, expressed as a deterministic sequence, 
is randomly selected according to a particular probability distribution,
This distribution is called an {\it information source} (or simply {\it source}), 
as every signal has a physical phenomenon as its origin.
Hence,  an observed value (deterministic sequence) is 
inherently associated with an underlying source (probability distribution)
\cite{Okamoto22}.

It is well known that certain patterns of the genome (e.g., non-coding DNA)
appear randomly distributed across a collection of many samples.
In such cases,
we can assume a source (probability distribution)
from which each specific genome pattern 
is randomly selected.
Here,, the actual object of complexity 
should not be viewed as each individual genome pattern,
but rather as the source (probability distribution) of those patterns.

Consequently, when representing an object of complexity 
as the outcome of its observation,
we should represent it by its source  (probability distribution) 
rather than by a specific observed value (deterministic sequence) 
randomly drawn from that source.

To represent an object as the source of an observed value,
we must employ an observation system 
For a given object, this system outputs 
an observed value (a deterministic sequence) along with 
its corresponding source (a probability distribution).
While the observed value is directly measurable, the source, though not directly visible, is inherently determined by the observation process when it produces the observed value.
Identifying the source from observed data is a classic problem
extensively studied within the framework of model selection theory in information theory and statistics. 
Notable examples include 
Minimum Description Length (MDL) \cite{Rissanen84,Rissanen89} 
and Akaike Information Criterion (AIC) \cite{Akaike73}. 
The objective is to identify the most probable and concise model that best represents the collected data.
While this paper does not focus on the methods of source identification,  
we assume that the source of an object's observed value is implicitly provided by the observation system. 

\begin{definition}[Object and Observation System]
\label{def:observation}
Let $A$ be an object, ${\cal O}$ be an observation system,
$c$ be a configuration (location and arrangement etc) of ${\cal O}$,
and ${\cal O}_{c}$ be observation system 
${\cal O}$ with configuration $c$.
For object $A$, the observation system ${\cal{O}}_{c}$ 
outputs ${\cal{O}}_{c}(A)$,
which consists of the observed value (deterministic), $\bar{\cal{O}}_{c}(A)$,
and the source (distribution) of the observed value,
$\hat{\cal{O}}_{c}(A)$. Thus, 
${\cal{O}}_{c}(A) \set (\bar{\cal{O}}_{c}(A), \hat{\cal{O}}_{c}(A))$.

Without loss of generality, $\bar{\cal{O}}_{c}(A)$ 
can be expressed in binary form 
because any physically observed data are bounded and have only finite precision. 
That is, we consider $\bar{\cal{O}}_{c}(A) \in \{0,1\}^n$ for some $n \in \N$, where
$\hat{\cal{O}}_{c}(A)$ is a probability distribution over $\{0,1\}^n$.
\end{definition}

A typical observed value consists of a deterministic component accompanied by noise.
Its source is the combined distribution of this deterministic value and 
the random distribution of the noise.
Note that a deterministic value can be considered a special  
case of a probability distribution 
(where only one value occurs with probability 1 and all others 
with probability 0)
when it is not just an observed value but 
the source of an observed value.

For many objects such as cosmic entities, genome and living organisms, 
the source of the observed value of an object %of many systems 
is the combined distribution of a deterministic part and 
a random part arising from noise or inherent randomness 
(e.g., non-coding DNA or internal fluctuations).
If an object (such as a galaxy, a living organism, or an advanced artificial construct) 
possesses a complex structure with %consisting of many components with
hierarchical and functional relationships,
the source of the observed value %of the object 
can be represented as the combined distribution of a deterministic part 
expressing that structure and 
a random part caused by noise or inherent randomness.
This is possible because such structures can be described deterministically.

Henceforth,
we represent any object as the source of its observed value,  
which is defined as a probability distribution.

%\paragraph{Complexity definition}
\subsubsection{Complexity Definition}
\label{sec:complexity-definition}

There are two types of complexity: disorganized and
organized \cite{Weaver48,Okamoto22}.
Disorganized complexity is characterized by randomness 
and is exemplified by %the complexity of 
systems with %composed of many 
numerous elements haphazardly distributed. %in a helter-skelter manner. 
Entropy serves as the quantitative measure of disorganized complexity.
In contrast, organized complexity is exemplified by 
systems such as stars, planets, galaxies, living organisms, the biosphere, 
and human societies. 
In these systems, elements are interconnected, %within a system, %an entity, 
forming a structured, organic whole through their mutual relations. 
Clearly, the complexity associated with the assumption presented %described 
in Section %I.B 
\ref{sec:arrow}
is organized complexity.

To define the organized complexity of a probability distribution,  
%of the observed data, 
such as %for example, 
the source of observed values, %data, %$\hat{\cal{O}}_{D,L}(A)$, 
we employ %utilize 
a quantitative definition of the complexity of organized matters,
termed {\it organized complexity} (OC) \cite{Okamoto22}.
Broadly, the OC for a probability distribution $X$ is defined
by the minimum size of a stochastic automaton form of circuit, the oc-circuit, 
required to simulate the distribution $X$.
In other words, the OC of $X$ is the length of the shortest description 
necessary to simulate $X$. 
(Refer to Appendix \ref{sec:oc} 
for an informal explanation of OC.) 
This definition is well-suited for our purposes 
as it is defined for probability distributions (the source of observed data)  
and fulfills most criteria 
for the quantitative measure of organized complexity  
\cite{Okamoto22} (Appendix \ref{sec:oc}). 

\begin{definition}[Organized Complexity (OC)]
Let $X$ be a probability distribution over 
$\{0,1\}^n$ %$n$ bit strings 
(for some $n \in \N$).
Then,
${\sf OC}(X, \delta)$ is the organized complexity (OC) of 
$X$ at a precision level $\delta$ ($0 \leq \delta < 1$) 
\cite{Okamoto22}(Appendix \ref{sec:oc}).  
\end{definition}

If $X$ is a simple random distribution over $\{0,1\}^n$,
which serves as the source of a random $n$-bit sequence (observed value),  
its OC is minimal (approximately $\log{n}$ bits) 
because its simulation description is very simple \cite{Okamoto22}. 
This is consistent with the fact that $X$ possesses 
very little organized complexity.
This also demonstrates why the input to OC must be 
the source rather than the observed value: 
the OC of an observed value (a deterministic sequence) is maximal ($n$ bits), 
as it requires a $n$-bit description to simulate.
Therefore, applying OC to the source of an object's observed value
%the methodology, OC for the source of the observed value of an object, 
correctly capture the object's inherent organized complexity. 

If $X$ is the source of a deterministic sequence accompanied by noise, 
it represents the combined distribution of the deterministic component and 
the random component caused by noise. 
In this case,
the OC of $X$ is principally determined by the deterministic part, 
as the OC of the random noise component is minimal.
This aligns with the intuition that the deterministic part contains 
the meaningful information, whereas the random part does not.
Furthermore, this confirms that the observed value is an inappropriate 
input to OC: the OC of an observed value cannot isolate the OC of the meaningful (deterministic) part because it is conflated with the OC of the noise sequence, which is almost maximal. 
Thus, applying OC to the source of an object effectively extracts the complexity of its meaningful components.

Note that when we heuristically separate a meaningful signal from noise 
in an observed sequence, we are implicitly applying the model selection theory described in Section \ref{sec:object} to identify the random noise component (corresponding to employing the source in our formulation) 
and exclude it  (corresponding to applying OC to it in ours).

If a source $X$ is the combined distribution of a deterministic part 
(expressing a complex structure) and 
a random part (caused by noise and inherent randomness),
the OC of $X$ is primarily derived from the deterministic part. 
Consequently,
this measure extracts the structural and organizational complexity of an object while filtering out its randomness or disorganized elements. 
This justifies our formulation of using  %our approach with using 
OC for the source of an observed value. 

\subsubsection{Collection of Observation Systems} 

Consider an observation system that monitors Earth's biosphere using a vast array of observation systems which are diverse instruments and devices. 
This system outputs a collection of observed values, each paired with its corresponding source. 
This collection consists of individual outputs from each instrument or device, rather than data centrally aggregated data within the system. 
In other words, the observation system lacks internal data transmission.

\begin{definition}[Collection of Observation Systems]
Let ${\cal O}$ be a collection of observation systems 
consisting of many observation systems which are
diverse instruments and devices, 
$(e_1, \ldots, e_m)$.
Let $c \set (c_1, \ldots, c_m)$ be a configuration 
of $(e_1, \ldots, e_m)$, 
where $c_i$ is a configuration  
of $e_i$ ($i=1,\ldots,m$).
Let ${\cal O}_{c} = ({e_1}_{c_1}, \ldots, {e_m}_{c_m})$  be 
the collection %of observation systems 
${\cal O}$ with configuration $c$, where 
${e_i}_{c_i}$ is $e_i$ with configuration $c_i$  ($i=1,\ldots,m$).
We assume these instruments and devices equally work in any location or environment. % location.

Let $A$ be an object.  
Let 
${\cal{O}}_{c}(A) = 
(\bar{\cal{O}}_{c}(A), \hat{\cal{O}}_{c}(A))$ %\in \{0,1\}^n$ ($n \in \N$) 
be the output of ${\cal O}_{c}$ observing $A$, 
where $\bar{\cal{O}}_{c}(A) = 
(\overline{e_1}_{c_1}(A), \ldots, \overline{e_m}_{c_m}(A))$ is the observed value of $A$ by 
${\cal O}_{c}$
and $\hat{\cal{O}}_{c}(A) = (\widehat{e_1}_{c_1}(A), \ldots, \widehat{e_m}_{c_m}(A))$ is the source (distribution) of 
$\bar{\cal{O}}_{c}(A)$,
that is, $\bar{\cal{O}}_{c}(A) \rset \hat{\cal{O}}_{c}(A)$, and 
$\overline{e_i}_{c_i}(A) \rset \widehat{e_i}_{c_i}(A)$  ($i=1,\ldots,m$).
Here, ${e_i}_{c_i}$ observes a part of $A$,
and ${e_i}_{c_i}(A)$ is the resulting output.
% of ${e_i}_{c_i}$ observing the part of $A$.

\end{definition}

As described in Section \ref{sec:object}, 
we consider $\bar{\cal{O}}_{c}(A) \in \{0,1\}^n$ for some $n \in \N$,
where $\hat{\cal{O}}_{c}(A)$ is a probability distribution over $\{0,1\}^n$.
 
When a collection of observation systems ${\cal O}$ observes $A$  
using various instruments and devices, $(e_1, \ldots, e_m)$, 
with configuration 
$c \set (c_1, \ldots, c_m)$
in $A$,
$\hat{\cal{O}}_{c}(A)$ represents
the tuple of the source of the observed values of $A$ produced  by 
$(e_1, \ldots, e_m)$,  
$(\widehat{e_1}_{c_1}(A), \ldots, \widehat{e_m}_{c_m}(A))$.
If the observation areas of these instruments and devices overlap, 
their outputs, $\hat{\cal{O}}_{c}(A)$, 
are expected to be redundant and interrelated. 
However, such redundancy %and overlap 
in $\hat{\cal{O}}_{c}(A)$ 
is eliminated in 
$ {\sf OC}(\hat{\cal{O}}_{c}(A), \delta)$,
the organized complexity (OC) of $\hat{\cal{O}}_{c}(A)$,
which is defined as the size of
the shortest description for simulating 
$\hat{\cal{O}}_{c}(A)$.
Therefore, we disregard the overlap of 
the observation areas of the devices in observation system $\cal{O}$
when considering the organized complexity, $ {\sf OC}(\hat{\cal{O}}_{c}(A), \delta)$.

For example, a vast number of observations have been made 
regarding the biological and geological phenomena on Earth. 
These observation systems have produced an enormous collection of observed values that are complexly overlapped and interrelated.
Such redundancy have been eliminated by  
unified theories and reports that explain these overlapping %observed 
values.
Following Occam's razor,
scientists strive to create the most concise theories possible 
by carefully removing noise and inherent randomness,
Hence, these theories correspond to OC's shortest description for simulating the source of the observations.

Even though the total volume of such observed data 
accumulated on the Internet and in databases
in the world is enormous, 
it remains finite and theoretically possesses a finite organized complexity value. 

\begin{definition}[Optimal Configuration]
Given an object $A$, 
a collection of observation systems ${\cal O}$, and a precision level $\delta>0$,  
the value of $ {\sf OC}(\hat{\cal{O}}_{c}(A), \delta)$ %varies 
depends on %by the 
the configuration $c$. % of ${\cal O}$. % \set (c_1, \ldots, c_m)$.
Let $c^*$ %_{A, D, \delta}  
be an optimal configuration of $c$ for $({\cal O}, A, \delta)$ such that 
$$
{\sf OC}(\hat{\cal{O}}_{c^*}(A), \delta)
= 
\max_c\{ {\sf OC}(\hat{\cal{O}}_{c}(A), \delta)\}.
$$
We then define
$$
{\sf OC}^*_{\cal{O}, \delta}(A) \set {\sf OC}(\hat{\cal{O}}_{c^*}(A), \delta).
$$

\end{definition}

\begin{claim}
\label{claim:optimal-configuration}
Given an object $A$, 
a collection of observation systems ${\cal O}$, 
and a precision level $\delta>0$,  
there exists 
an optimal configuration of $c^*$ for $({\cal O}, A, \delta)$. 
\end{claim}

%\vspace{2pt}
\noindent
{\it Proof}

Since $\delta>0$ (finite precision),   
the value (bit length) of ${\sf OC}(\hat{\cal{O}}_{c}(A), \delta)$
is a finite natural number for any $c$.
Hence, the number of variations of $c$ 
that effectively alter the value of ${\sf OC}(\hat{\cal{O}}_{c}(A), \delta)$
is also finite, given $(\cal{O}, A, \delta)$.
By exhaustively searching the finite number of the effective variations of $c$,
an optimal configuration $c^*$ can be found such that 
$$
{\sf OC}(\hat{\cal{O}}_{c^*}(A), \delta)
= 
\max_c\{ {\sf OC}(\hat{\cal{O}}_{c}(A), \delta)\}. 
\hspace{70pt}
\square
$$

For instance,
let $A$ represent the Earth's biosphere, and consider 
a collection of observation systems, $\cal{O}$, 
consisting of a vast number of instruments and devices to observe the biosphere.
While there are countless possible configurations 
(e.g., locations and arrangements), $c$,
%of the instruments and devices within the biosphere, 
Claim \ref{claim:optimal-configuration} ensures
there exists an optimal configuration $c^*$
that maximizes the OC of the source of 
the observed values of $A$ by $\cal{O}$. 
In practice, we empirically determine the best possible configuration 
to maximize the amount of meaningful information captured. 

We then introduce an order relation between two observation systems.

\begin{definition}[Order Relation] % for Observation Systems]
Let ${\cal{O}}$ and ${\cal O}'$ be two collections of observation systems.
We denote ${\cal O} \geq{\cal O}'$ iff
for any object $A$ and any precision $\delta > 0$,
${\sf OC}^*_{\cal{O}, \delta}(A) \geq {\sf OC}^*_{\cal{O}',\delta}(A)$.
\end{definition}

For example,
let
${\cal O} \set (e_1, \ldots, e_m)$ and ${\cal O}' \set (e_1, \ldots, e_\ell)$,
where $m> \ell$,
that is, $\cal{O}$ contains all instruments of $\cal{O}'$ plus additional ones,
$(e_{\ell+1}, \ldots, e_m)$. And, 
$\bar{\cal{O}}_{c}(A) \in \{0,1\}^n$ for some $n \in \N$,
and $\bar{\cal{O}'}_{c'}(A) \in \{0,1\}^{n'}$ for $n' < n$.
Then, for any object $A$ and any precision $\delta  > 0$,
${\sf OC}^*_{\cal{O}, \delta}(A) \geq {\sf OC}^*_{\cal{O}',\delta}(A)$.
This is because:
Let ${c'}^*$ be the optimal configuration for $({\cal O}', A, \delta)$, and
configuration of $c$ for $(\cal{O}, A, \delta)$ be the same as 
${c'}^*$ for $(e_1, \ldots, e_\ell)$ and arbitrary for $(e_{\ell+1}, \ldots, e_m)$.
Then,
${\sf OC}^*_{\cal{O}, \delta}(A) 
\geq {\sf OC}(\hat{\cal{O}}_{c}(A), \delta)
\geq  {\sf OC}^*_{\cal{O}',\delta}(A)$.

The relation ${\cal O} \geq{\cal O}'$ indicates that
${\cal{O}}$  can capture any property of complexity that 
${\cal O}'$ captures.
This order relation is useful for demonstrating  
that the complexity of object $A$ possesses a generic property, %of complexity 
one that does not depend on %independent of
any specific observation system. 
%This is because 
The complexity of object $A$ can be considered to 
have such a generic property if there exists 
an observation system ${\cal O}'$ such that 
${\sf OC}^*_{{\cal O}, \delta}(A)$ satisfies %possesses %
the property for all ${\cal O} \geq{\cal O}'$.

\subsection{Macroscopic Phenomena of Increasing Complexity}
\label{sec:increasing-complexity-phenomena}

In this section, we roughly overview the phenomena of increasing organized 
complexity in %non-equilibrium systems,
the universe and the Earth's biosphere.
We employ heuristic approximations of organized complexity
as the (shortest) description length required to characterize 
the structural and organizational features of an object 
(Section \ref{sec:complexity-definition}).
These approximations are derived from intuitive assessment, 
which is justified here 
because we primarily addresses relative scales and temporal fluctuations (increases or decreases) in organized complexity.

It is thought that immediately after the universe was born 13.8 billion years ago, it existed in a state of extremely high temperature with no elementary particles. Its organized complexity at that time could be said to 
have been at its lowest level. 

Subsequently, elementary particles formed, followed by light atoms such as  hydrogen and helium. 
Compared to a universe with no elementary particles, the emergence of particles and light atoms increased the organized complexity of the universe by increasing its necessary description length.  

Over time, numerous protostars were born.
Stellar activity and supernova explosions created a variety of heavy atoms, which chemically combined to form diverse molecules. This led to the birth of planets, planetary systems, and stars containing heavy atoms; eventually, structures such as galaxies, galaxy clusters, and superclusters emerged. Describing such a universe requires a significantly greater description length, indicating that the universe achieved higher organized complexity.

Earth formed in the solar system approximately 4.6 billion years ago. The first life forms, prokaryotes, are believed to have emerged about 4 billion years ago, followed by eukaryotes (single-celled organisms) around 2 billion years ago. While the biosphere from 4 billion to 2 billion years ago consisted solely of prokaryotes, the subsequent inclusion of eukaryotes increased the organized complexity of the biosphere.

Thereafter, the variety or the organized complexity of the biosphere increased sequentially. For instance, multicellular organisms emerged approximately 1 billion years ago, arthropods around 540 million years ago, and mollusks (such as cephalopods) around 490 million years ago. New types of organisms were added to the biosphere in succession; today, the biosphere consists of a vast array of diverse species.

Consequently, the organized complexity of the biosphere, including humans, is thought to have increased over time. While complexity decreased temporarily during major mass extinction events, it subsequently recovered and increased beyond its previous levels.

\subsection{Macroscopic Law and Hypothesis}
\label{sec:law-hypothesis}

We now describe the {\it law of increasing complexity} 
as a macroscopic (emergent) law in the universe.
This law formulates the assumption that 
the universe in general, and the biosphere on Earth in particular, is becoming more complex over time (Section \ref{sec:arrow})
as well as the observation in Section \ref{sec:increasing-complexity-phenomena},
using the concept of the organized complexity (OC) 
\cite{Okamoto22} (Appendix A) as applied to
the source of an object's observed value
(Section \ref{sec:complexity-definition-any-objects}).

\vspace{5pt}
\noindent
{\rm [}{\bf Law of Increasing Complexity}{\rm ]}

{\it 
Let $\SSS$ be a system consisting of numerous elements. 
The state of $\SSS$ changes over time, and let $\SSS_T$ denote the state of $\SSS$ at time $T$, where time in $\SSS$ is determined by an inertial frame of reference in $\SSS$ (e.g., on Earth).

We say that the law of increasing complexity emerges in 
system $\SSS$ if the following holds:
 
System $\SSS$ is associated with  
a time interval $I_\SSS \set [T_0, T_1]$ ($T_0, T_1 \in \R^+$, $T_0 < T_1$), 
an observation system 
${\cal O}_{\SSS}$, and a precision level $\delta_\SSS$.

\noindent
Then, for any collection of observation systems ${\cal O} \geq {\cal O}_{\SSS}$, any precision level $\delta$ ($0< \delta \leq \delta_\SSS$), 
and 
any time $t \in I_\SSS$, 
there exists $s \in \R^+$ such that
$$
{\sf OC}^*_{{\cal O},\delta}(\SSS_{t+s})
>  {\sf OC}^*_{{\cal O},\delta}(\SSS_t).
$$
}

%\vspace{5pt}
\noindent
{\bf Remark 1} \ 
This law does not imply that complexity 
increases monotonically 
(i.e., it does not claim that for any $t$, \ 
for any $s>0$, \ 
${\sf OC}^*_{{\cal O},\delta}(\SSS_{t+s})
>  {\sf OC}^*_{{\cal O},\delta}(\SSS_t)
$).
Instead, it asserts that while complexity may temporarily decrease, it will inevitably grow to exceed previous levels at some later time
(i.e., for any $t$, \ there exists $s>0$ such that
$
{\sf OC}^*_{{\cal O},\delta}(\SSS_{t+s})
>  {\sf OC}^*_{{\cal O},\delta}(\SSS_t)
$).
For example, in the biosphere, complexity has occasionally decreased due to mass extinctions, but it subsequently rose beyond all prior levels.

%\vspace{5pt}
\noindent
{\bf Remark 2} \ 
As defined in Section \ref{sec:notation},
a probability distribution $\hat{\cal{O}}_{c^*}(A)$ over $\{0,1\}^n$,  representing the source of  the ${\cal O}_{c^*}$'s observed value of 
object $A$, %for OC, 
is expressed as
$\{ (a, p_a) \mid a \in \{0,1\}^n, p_a \in [0,1], 
\sum_{a\in \{0,1\}^n}p_a=1 \}$. 
Since an oc-circuit for % OC 
$
{\sf OC}^*_{\cal{O}, \delta}(A) \set {\sf OC}(\hat{\cal{O}}_{c^*}(A), \delta)
$
employs a random source of finite length
($(r_1, \ldots, r_K)$ in Appendix \ref{sec:oc}),
it outputs a simulated distribution for $\hat{\cal{O}}_{c^*}(A)$ as 
$\{ (a, p'_a) \mid a \in \{0,1\}^n, p'_a \in [0,1], 
\sum_{a\in \{0,1\}^n}p'_a=1 \}$
where
$p'_a$ is a ``discrete value'' 
(whereas, $p_a \in \R$).
Consequently, there exists a statistical distance 
(representing a precision level) 
$\delta = \frac{1}{2}\cdot 
\sum_{a\in \{0,1\}^n} | p_a - p'_a| \geq 0$.
The smaller the $\delta$ (i.e., the higher the precision),  
the greater randomness $(r_1, \ldots, r_K)$
is required for the oc-circuit, 
resulting in a higher value of ${\sf OC}^*_{\cal{O}, \delta}(A)$ 
(a larger size of circuit $\overline{C}$ in Appendix \ref{sec:oc}).
The value of $\delta_\SSS$ represents the minimum precision level
necessary to recognize the phenomena of increasing complexity.

%\vspace{5pt}

%We then present a hypothesis regarding this macroscopic law.

\vspace{5pt}
\noindent
{\rm [}{\bf Hypothesis of Increasing Complexity}{\rm ]} 

{\it  
There exist 
non-equilibrium systems with abundant free energy flows
in the universe %for 
in which 
the law of increasing complexity has emerged. 
Specifically, 
this law has emerged in the following 
systems:  
}
\begin{itemize}
\item
{\it The observable universe}:  
A non-equilibrium system with abundant free energy flows
comprising all matter within the observable universe as seen from Earth,  including the Earth's biosphere.
The observable universe is the spherical region centered on Earth with a radius of approximately 46.5 billion light-years (a comoving distance that accounts for the expansion of the universe since the Big Bang).

\item
{\it The biosphere on Earth}:  
A non-equilibrium system with abundant free energy flows
comprising all living organisms (including humans) and their associated environments on Earth. 

\end{itemize} 

%\vspace{5pt}
\noindent
{\bf Remark 3} \ 
For a system $\SSS$ under this hypothesis,   
the time interval $I_\SSS$ spans from its inception (e.g., the birth of  
$\SSS$) to certain point in the future.
For the Earth's biosphere, the interval extends from the origin of life to a point preceding its eventual extinction, potentially caused by increasing 
solar luminosity.
The end of the interval for the observable universe remains uncertain 
but is expected to arrive in the distant future, as  
the entropy of the entire universe increases in accordance with 
the second law of thermodynamics (see 
Section \ref{sec:future} for further discussion).   

%\vspace{5pt}
\noindent
{\bf Remark 4} \ 
Direct verification of the law of increasing complexity in the Earth's biosphere and the observable universe is challenging due to the inherent limitations of observing past events directly. However, we can indirectly infer past observations. For the biosphere, we infer them %past observations 
through the analysis of fossils, genetic data, and historical records. 
For the observable universe, we can observe past states by examining astronomical objects at varying distances; due to the finite speed of light, observing distant objects is equivalent to observing the universe as it existed in the past. We can refine our inference of the cosmic past observation by leveraging the vast wealth of data collected by advanced astronomical instruments.

%\vspace{5pt}
\noindent
{\bf Remark 5} \ 
While the observable universe and the Earth's biosphere are immense from our perspective, they are theoretically finite. They can be measured by a collection of observation systems at a certain precision level, provided the collection includes a sufficiently large number of diverse observation systems. Regarding the Earth's biosphere, a vast network of observation systems has already been established, generating an enormous volume of data. While it is challenging to evaluate the exact OC of the source of this observed data, it is theoretically possible. It should be noted that this law and the associated hypothesis are formal theoretical statements (see Section \ref{sec:reduction} for their theoretical significance).  As discussed in Section 
\ref{sec:increasing-complexity-phenomena}, we can obtain heuristic approximations through intuitive assessment to represent the relative scales and temporal fluctuations of organized complexity in practice.

\subsection{Significance and Reduction of This Law} 
\label{sec:reduction}

There are two primary reasons why formulating the law of increasing complexity is significant.
First, this law allows us to effectively explain related macroscopic phenomena 
in a manner similar to that heat phenomena are 
effectively explained by the macroscopic laws of thermodynamics rather than 
fundamental physical laws.

Second, the formulation of this law provides a theoretical foundation for the phenomena of increasing complexity  
through its potential reduction to fundamental physical laws.  
In the scientific tradition, 
a physical phenomenon is considered fully understood 
only when it is explained by, or reduced to the fundamental physical laws. 
To proceed with this reductionist approach, the target phenomenon must 
first be formally defined.
Our paper establishes this necessary stage for the phenomena of increasing complexity by formulating it as the law of increasing complexity.
%\end{itemize}

To complete this reductionist approach, a critical challenge remains for the next stage: the (approximate) reduction of this law to fundamental physical laws. Such a reduction would provide the ultimate certification for the law of increasing complexity.

Previous studies on dissipative structures and self-organizations have 
explored how order emerges in natural environments \cite{NicPri77,Kauffman93}.
Furthermore, the free energy principle proposed by Friston explains 
brain function, perception, and action 
as a self-organizational process aimed at minimizing 
a system's variational free energy (a value related to uncertainty and prediction error) \cite{Friston10}.
These existing frameworks may help identify a unifying principle for the emergent phenomenon of increasing complexity, ultimately leading to the reduction of this macroscopic law to fundamental physics.

\subsection{Two Modes of Increasing Complexity}

Because the law of increasing complexity may emerge in numerous systems within the universe, 
it suggests two potential scenarios as described in Sections \ref{sec:issues}:
(1) The average (or total) complexity of the (observable) universe is increasing, and
(2) The complexity of the most (or highly) complex systems within the universe is increasing.

The hypothesis of increasing complexity (Section \ref{sec:law-hypothesis})
posits that both scenarios are occurring simultaneously. 
That is, 
(1) The total complexity of the observable universe is increasing. 
(2) The complexity of a very highly complex system within the universe,
the Earth's biosphere, is also increasing.

\subsection{Consistency with the Second Law of Thermodynamics}

If the macroscopic law of increasing complexity 
emerges in certain systems in the universe,
it is not 
inherently inconsistent with   
the second law of thermodynamics or the 
law of increasing entropy,  
since both macroscopic laws %ultimately 
emerge from the same fundamental physical laws governing the universe;  they describe different aspects of the arrow of time under differing conditions (e.g., isolated vs. non-equilibrium systems).

\subsection{The Hypothesis in the Distant Future}
\label{sec:future}

The available free energy ${\Delta}S$, defined as
the entropy gap between the maximum possible entropy 
and the actual entropy of the universe, represents
the source of free energy flows. 
The combination of the cosmological constant and the second law of thermodynamics requires that ${\Delta}S \ra 0$
%the available free energy is approaching zero
(as discussed in Chapter 3 of \cite{LinDavRus13}).

Consequently, 
non-equilibrium systems with abundant free energy flows 
will vanish in the very distant future. % in the universe.
This suggests that the hypothesis of increasing complexity
will eventually fade away, %in the very distant future,
and the organized complexity of the universe will approach zero
as the cosmos nears heat death, or the state where ${\Delta}S =  0$.
  
\subsection{Physical Laws and Computation Theory}

Mathematics serves as the language for describing physical laws and theories. 
Various mathematical frameworks, including entirely novel approaches, 
have been developed to provide the necessary tools for this description.
It may seem unusual that the law of increasing complexity 
presented here utilizes a computational concept, 
the {\it oc-circuit}, to address
physical phenomena. 
However, computation theory is a rigorous mathematical discipline
founded in the 1930s by mathematicians such as Turing and 
G{\"o}del \cite{Sipser96}. 
Just as calculus or differential geometry are employed 
in other areas of physics,
computation theory provides a suitable and precise framework  
for describing its emergent macroscopic law.

\subsection{Related Works}

While this paper formulates the law of increasing complexity through a universal methodology, OC for the source of observed values, to measure the complexity of {\it any observable object}, other studies have explored increasing complexity within specific domains. Notable examples include 
{\it functional information} for the complexity of certain functions
\cite{HazGriCarSzo07,HazWon24} 
and {\it assembly theory} for the complexity of molecules 
\cite{Marshall-etal21}.

%\paragraph{Functional Information (FI)}

The functional information (FI) of function $x$,
(e.g., a folded RNA sequence that binds to GTP)
within a system is defined by its complexity, 
$-\log_2{F(E_x)}$. Here $E_x$ represents the degree of function of $x$
(e.g., the RNA–GTP binding energy), and 
$F(E_x)$ is the fraction of all
possible configurations of the system that possess a degree of
function $\geq E_x$. 

For a system with $N$ possible configurations 
(e.g., a sequence of $n$ RNA nucleotides, 
which has $N=4^{n}$ discrete possible sequences),
the FI of $x$ is given by $-\log_2{M(E_x)/N}$,
where $M(E_x)$ is the number of configurations 
whose degree of function meets or exceeds
the specified degree of function of $x$, $E_x$.
According to information theory
\cite{CovTho06},
the value of $-\log_2{M(E_x)/N}$
represents the minimum information 
(the shortest description size) required to
identify the functional feature of $x$ shared by
$M(E_x)$ configurations among the total $N$.  

If the features of function $x$ are observed,
the OC of the source of the observed value,
which is the shortest description size %(by oc-circuit)
for simulating of the source or the features of function $x$,
could be consistent with the FI of function $x$.
In this sense,
FI may provide a roughly approximation of the OC for the source of 
observed functional features. 
 
%\paragraph{Assembly Theory (MA)} 
   
In assembly theory,
the molecular assembly index (MA) is used to quantify 
the complexity of molecules.
The MA of an object is defined as the length of the shortest 
assembly pathway; that is, the minimum number of joining operations
required to construct the object, where intermediate objects created 
during the process can be reused. 
Assembly pathways are sequences of joining operations, 
that start with basic building blocks and terminate with a final product.

The authors of \cite{Marshall-etal21} developed an analytical method to correlate experimental mass spectrometry data directly with MA. 
Because MA is closely related to the structural heterogeneity of molecules, 
it serves as a robust biosignature.
Since the MA of a molecule represents its shortest assembly pathway,
it is inherently related to the molecule's shortest description length.
If the molecule is analyzed via mass spectrometry,
the OC of the source of the observed value,
the shortest description length 
for simulating of the source, 
could be consistent with its MA.
Thus,
the MA of a molecule or its mass spectrometry data 
may provide a viable %roughly 
approximation of our OC measure.

Although our methodology can rigorously define the (organized) complexity of any observable object in theory, measuring these values in practice remains a significant challenge. Here note that, however, as discussed in 
Section \ref{sec:reduction}, this formulation provides a theoretical foundation for the phenomena of increasing complexity. 
%through its potential reduction to fundamental physical laws. 
Identifying practical methods for approximating these values is highly valuable. Both functional information and assembly theory offer concrete ways to measure the complexity of specific objects. These measures appear consistent with our methodology and represent promising proxies for obtaining approximate values of organized complexity.
 
\section{An Application of This Law to the Fine-Tuning Problem}

\subsection{Fine-Tuning Problem and Explanations}
\label{sec:coincidence}

The fundamental physical laws incorporate various physical constants, including  the gravitational constant, the speed of light, Planck's constant, and the elementary charge. 
However, there is as yet no definitive explanation for why these constants take on their currently measured values.
At present, it might be said that these values are a result of chance, though a future “theory of everything” may eventually provide an answer.

One remarkable observation is that the physical constants of our universe appear to be set to values conducive to the creation of life on Earth in such an intricately balanced way that they could hardly be a product of pure chance
\cite{Friederich21,Davies07}. 
For example, the strong force determines the magnitude of the nuclear force binding elementary particles in the atomic nucleus, while the electromagnetic force determines the magnitude of the force binding electrons to the nucleus and creating repulsion between protons.
If the balance between these two constants were to deviate even slightly,
for instance, if the electromagnetic force were slightly stronger, atoms other than hydrogen could not exist, meaning no carbon and, by extension, no life on Earth. Conversely, if the strong force were slightly stronger, hydrogen would disappear, likewise precluding the existence of life. Numerous other examples of physical constants appearing fine-tuned for life are well-documented
\cite{Friederich21,Davies07,SmiSmiVer18,Davies03}.
The {\it fine-tuning problem} questions why the 
physical constants appear to have been fine-tuned for life on Earth. 

Various explanations for this problem exist, utilizing the anthropic principle
 \cite{Davies07,BarTip86}  
 or extra-universal hypotheses such as the multiverse
\cite{Carr07}, 
{top-down cosmology} \cite{HawHer06}, and
{cosmic designers}  \cite{Friederich21,Dick15,Mizrahi17}.

The weak anthropic principle states that 
the observed physical laws and constants %in the universe 
must be consistent with the existence of the observer, such as (human) life.
It explains the fine-tuning problem by asserting that   
the physical constants we observed must %should 
be fine-tuned for us to be here to observe them.
However, this is often criticized as a tautology 
concerning observation selection effects.

The multiverse hypothesis posits %proposes 
the existence of numerous universes,
each with different values of fundamental physical constants. 
It suggests that our universe is simply one of the rare instances where the constants happened to be appropriate for life. 
This invokes selection bias to support the weak anthropic principle.

The top-down cosmology proposes 
that the universe's initial conditions were  %consisted of 
a superposition of many possibilities, 
only a tiny fraction of which contributed to the universe we observe today.
This offers a weak anthropic explanation rooted in quantum cosmology.

Variants of cosmic designers hypothesis include
extra-universal aliens or intelligent entities who designed the universe,
program simulators of the universe, and
a supernatural agent or God. 
These explanations %for the fine-tuning problem from the cosmic designers     
rely on %would be 
teleological arguments similar to the strong anthropic principle.

In summary, 
the weak anthropic principle is tautological, while the multiverse and top-down cosmology support it by invoking selection bias.
The cosmic designers hypothesis suggests a version of the strong 
anthropic principle through teleological reasoning.
Consequently, almost all current explanations are linked to the anthropic principle because they assume the universe is fine-tuned for ``life'' on Earth.

However, there is significant debate regarding whether "life" is the true objective of this fine-tuning \cite{Friederich21}.
Many physicists and cosmologists suggest that 
the universe is fine-tuned not so much for "life" itself, but rather for the ``building blocks and environments that life requires''
\cite{SmiSmiVer18,Davies03}.
A caution is raised that the range of parameters for which the universe would have been habitable is quite considerable \cite{Adams19}. 
Furthermore, the universe could have been more, rather than less, life-friendly
if some parameters of the physical laws were to change slightly.
Our universe may just be habitable rather than maximally life-supporting.
Even the weak anthropic principle states that 
this universe is consistent with life, but it does not mean that it is optimally fine-tuned for life.
%Hence, 
Our universe may be more fine-tuned for something else than life.
We may be biased to assume that all possible kinds of complex organizations 
in the universe will resemble life on Earth. 
The creation of life on Earth is just one phenomenon in the universe, and 
numerous other phenomena could occur, such as the formation of diverse complex organizations 
beyond our current imagination. 

This leads to a pivotal question:

%\vspace{5pt}
{\it Can we explain the fine-tuning problem 
using a more plausible objective 
than life or the anthropic principle?}

\subsection{New Explanation}

We address the question in Section \ref{sec:coincidence}
using the hypothesis introduced in Section \ref{sec:law-hypothesis}.

In this section, we refer to the observable universe simply as "the universe." Thus, our hypothesis can be succinctly stated: “The law of increasing complexity emerges in the universe.”

Macroscopic phenomena, such as the law of increasing complexity, 
emerge from fundamental physical laws. 
Therefore, this hypothesis implies that 
``the law of increasing complexity 
emerges from the fundamental physical laws
in this universe.''
If these fundamental physical laws were to change, 
any macroscopic phenomena and laws emergent from them would 
necessarily be affected.
Consequently, any significant alteration to the physical constants within 
these laws would likely prevent the emergence of the law of increasing complexity.

With this in mind, 
our new explanation of the fine-tuning problem %, based on this hypothesis, 
posits that
``the fundamental physical laws are configured so that the
law of increasing complexity emerges in this universe.''

In other words, 

\vspace{5pt}
\noindent
{\rm [}{\bf 
Our explanation of  the fine-tuning problem}{\rm ]}

{\it The fundamental physical constants 
are fine-tuned for the emergence of the %(macro) 
law of increasing complexity in the universe. }

%\vspace{5pt}

Let us revisit the example from Section \ref{sec:coincidence}
concerning the delicate balance between the nuclear and 
the electromagnetic forces.
If these constants were to deviate slightly, disrupting this delicate balance, 
for example,
if the electromagnetic force were to be slightly stronger, 
no atoms other than hydrogen could form.
This would preclude the formation of more complex structures, 
potentially leading to a chaotic state or a still universe resembling heat death. 
As a consequence, the law of increasing complexity would fail to emerge.

Conversely, if the strong force were to be slightly stronger, 
hydrogen would disappear, resulting in a universe dominated by heavier atoms. This would %significantly 
severely limit molecular diversity;
for instance, it would prevent the formation of water, methane, ammonia, and hydrides including essential acids like amino acids and nucleic acids, as well as various macromolecules. There would likely be far fewer stars and galaxies than we currently observe (given that roughly 75{\%} of the baryonic matter in the current universe is hydrogen by mass). Consequently, only a limited degree of order could arise, hindering the law of increasing complexity. 
Based on these observations, we hypothesize that the measured values of these constants are fine-tuned specifically for the emergence of this law.

In the light of the law of increasing complexity, 
we can consider two distinct types of universes
within the multiverse hypothesis.
%We now focus on two types of universes
%in light of the law of increasing complexity.
The first type is a universe where the law of increasing complexity emerges, characterized by richness, diversity, and an abundance of order.
Here, this law emerges in the sense of the hypothesis in Section \ref{sec:law-hypothesis}.
That is,
in the first type of a universe,
there exist non-equilibrium systems with abundant free energy flows
where the law of increasing complexity has emerged.
Our universe exemplifies this type. 
The second type is a universe where the law of increasing complexity does not emerge, resulting in limited order and structure. Such universes might remain in a chaotic state or reach a state of equilibrium (heat death) very quickly.

Note that the law of increasing complexity assumes the existence of
observation systems (per Definition \ref{def:observation}). 
Hence, if this law emerges in a universe,
observation systems exist under its physical law. 
Conversely, observation systems do not necessarily exist
in a universe where this law fails to emerge. 

\subsection{Advantage}

Our explanation is grounded in an objectively formulated emergent law, 
the law of increasing complexity, which utilizes the mathematically defined notion of organized complexity (OC). It does not rely on subjective or anthropocentric ideas such as the anthropic principle or teleological arguments.

This explanation suggests a natural emergent relationship between fundamental and macroscopic laws. The emergence of life is just one phenomenon among many; the emergence of this macroscopic law represents a more fundamental and profound property of the universe, paving the way for life, its required environments, and a diverse array of complex organizations. It is more plausible that the universe is fine-tuned for a fundamental property rather than a specific, local phenomenon like life.

If we could achieve an (approximate) reduction of this macroscopic law to fundamental physical laws (as discussed in Section \ref{sec:reduction}), it would allow us to concretely evaluate the conditions required for this law to emerge. Such a reduction would clarify whether the physical constants are optimally fine-tuned for this objective.

Under the multiverse hypothesis, if the physical constants of a newly formed universe were randomly determined, the law of increasing complexity would likely fail to emerge in the vast majority of cases. A universe where this law does emerge, resulting in a rich, varied, and ordered cosmos, would be rare. In such rare universes, the physical constants are fine-tuned for the law, exactly as we observe in our own.

\section{Conclusion}

This paper develops a novel methodology for the macroscopic law of increasing complexity, which formulates the arrow of time in terms of increasing organized complexity for certain systems. It introduces a framework for measuring the complexity of any observable object, providing a quantitative definition of organized complexity (OC) for the source of an object's observed value, alongside an order relation between observation systems. We hypothesize that this macroscopic law emerges in certain non-equilibrium systems with abundant free energy flows, such as the observable universe and the Earth's biosphere.

The formulation of this law provides a theoretical foundation for the phenomenon of increasing complexity through its potential reduction to fundamental physical laws. 

We apply this formulation to the fine-tuning problem of fundamental physical constants. Our new explanation posits that these constants, and the laws they inhabit, are fine-tuned for the emergence of the law of increasing complexity in the universe.

A critical remaining challenge is the (approximate) reduction of this macroscopic law to fundamental physical laws. Such a reduction would validate this hypothesis and allow for a concrete evaluation of the conditions on fundamental physical constants required to permit the emergence of this law.

\appendix

\section{Organized Complexity (OC)}
\label{sec:oc}
Here, we show the definition of the organized complexity (OC) roughly
(see \cite{Okamoto22} for the precise definitions and notations).

\begin{itemize}

\item
The {\it oc-circuit} (stochastic automaton form of circuit) is defined as below. 
Let $C$ be a circuit with $N$ input bits and $L$ output bits 
($C: \{0,1\}^N \rightarrow \{0,1\}^L$),
and $\overline{C}$ be a specified form of $C$.

$(\overline{C}, u, n, (m_1,..,m_K))$ is an oc-circuit ${\cal C}$   
 such that
\begin{eqnarray*}
&&
(s_{i+1}, y_{i}) \longleftarrow 
\fbox{$C(u, \ \cdot \ )$}
\longleftarrow (s_{i}, m_i, r_i), 
\\
&&
\ \ \ \ \ \ \ \ \ \ \ \ \ \ \ \ \  \ \ \ \ \ \ \ \ \ \ \ \ \ \  \ \ \ \ \ 
i=1,2,..., K, 
\\
&&
Y \set (y_1,...,y_K)_n,  
\ \ \ \ \ 
\end{eqnarray*}
where
$(y_1,...,y_K)_n$ is the $n$-bit prefix of $(y_1,...,y_K)$, 
$s_{i}$ is a state at step $i$, 
$u$ is an a priori input (universe),  
$m_i$ is an input at step $i$,
$r_i$ is random bits at step $i$,
$K \set \lceil n/L \rceil$, \ $L = |y_i|$ (bit length of $y_i$) for all $i$.

\item
{\it Organized complexity} $\OC$ of an object, 
distribution $X$ over $n$ bit strings, 
at precision level $\delta$ ($0\leq \delta < 1$) is defined as 
\begin{eqnarray*}
&&
\OC(X,\delta) := 
\min\{ |{\cal C}| \mid  X \scl  Y \rset {\cal C}\}, \ 
\end{eqnarray*}
where 
${\cal C}$ 
is an oc-circuit, and
$X \scl Y \rset {\cal C}$ 
means the output (simulation result) $Y$ of 
oc-circuit ${\cal C}$ is close (with precision level $\delta$) to object $X$
($|{\cal C}|$ denotes the bit length of ${\cal C}$).
That is, the organized complexity of object $X$
is the shortest size of oc-circuit ${\cal C}$ that simulates $X$ (with precision level $\delta$).  \ \ \

\item
{(Comparison)}

We give five criteria required for organized complexity definitions and
summarize the comparison of this definition (OC)
with other definitions 
in Table \ref{table:comparison} 
in terms of the criteria \cite{Okamoto22}.
As shown in Table \ref{table:comparison}, 
OC  satisfies all of the criteria.

\end{itemize}

%\vspace{500pt}
%\onecolumn

\begin{table*}[h]
\begin{center}
\caption{Comparison} % Regarding Criteria}
\label{table:comparison}
\begin{tabular}{|c||c|c|c|c|c|c|c|} 
\hline
Definitions of complexity&Objects&Simply&Simply
&\multicolumn{3}{c|}{Key features of highly organized objects} & Computa- \\ 
\cline{5-7}
& (Prob./Det.)&regular&random& Descriptional  & Computational & Distributional & bility\\
\hline\hline
{Kolmogorov complexity} &$\times$&{\checkmark}&$\times$&
{\checkmark}&$\times$&$\times$&$\times$ \\ \hline
Effective complexity &$\times$ &{\checkmark}&{\checkmark}&
{\checkmark}&$\times$&$\times$&$\times$ \\ \hline
Logical depth          &$\times$&{\checkmark}&{\checkmark}&
$\times$&{\checkmark}&$\times$&$\times$ \\ \hline
Effective measure complexity  &$\times$&{\checkmark}&{\checkmark}&
$\times$&$\times$&{\checkmark}&{\checkmark} \\ \hline
Statistical complexity      &$\times$&{\checkmark}&{\checkmark}&
$\times$&$\times$&{\checkmark}&{\checkmark} \\ \hline\hline
Organized Complexity (OC)      &{\checkmark}&{\checkmark}&{\checkmark}&{\checkmark}&{\checkmark}&{\checkmark}&{\checkmark} \\ \hline
\end{tabular}
\end{center}
\end{table*} 
 

\begin{thebibliography}{9}

\bibitem{Anderson72}
Anderson P. W. More Is Different, 
Science, Vol. 177, No. 4047, pp. 393-396  (1972).

\bibitem{Laughlin05}
Laughlin R. B.
A Different Universe: Reinventing Physics from the Bottom Down,
Basic Books (2005).


\bibitem{NicPri77}
Nicolis G. and Prigogine I.
Self-Organization in Nonequilibrium Systems: 
From Dissipative Structures to Order Through Fluctuations,
Wiley (1977).


\bibitem{LinDavRus13}
Lineweaver C. H.,  Davies P. C. W., and Ruse M. Eds. 
Complexity and the Arrow of Time, Cambridge University Press (2013).

\bibitem{Weaver48}
Weaver W. Science and Complexity, American Scientist 36 (4): 536--544 (1948).

\bibitem{Okamoto22}
Okamoto T. A New Quantitative Definition of the Complexity of Organized Matters,
Complexity, Volume 2022, Article ID 1889348 (2022)

https://www.hindawi.com/journals/complexity/2022/
1889348/


\bibitem{Rissanen84}
Rissanen J. Universal coding, information, prediction, and estimation,
IEEE Transactions on Information Theory, vol. 30, no. 4, pp. 629–636 (1984).

\bibitem{Rissanen89}
Rissanen J. Stochastic Complexity in Statistical Inquiry, 
World Scientific, Singapore (1989).

\bibitem{Akaike73}
Akaike H. Information Theory and an Extension of the Maximum Likelihood Principle, 
Proceedings of the 2nd International Symposium on Information Theory, 
Akadimiai Kiado, Budapest, Hungary (1973).


\bibitem{Kauffman93}
Kauffman S.A.
The Origins of Order: Self-Organization and Selection in Evolution,
Oxford University Press (1993). 

\bibitem{Friston10}
Friston K.
The free-energy principle: a unified brain theory?, 
Nature reviews. Neuroscience, 11(2), 127-38 (2010). 

\bibitem{Sipser96}
Sipser M.  Introduction to the Theory of Computation, Third Edition, Cengage Learning (2013).

\bibitem{HazGriCarSzo07}
Hazen R. M., Griffin P. L., Carothers J. M., and Szostak  J. W. Functional information and the emergence of biocomplexity, PNAS, vol. 104, suppl. 1, pp. 8574–8581 (2007). 
doi.org/10.1073/pnas.0701744104

\bibitem{HazWon24}
Hazen R. M., and Wong M. L. Open-ended versus bounded evolution: Mineral evolution as a case study, PNAS Nexus, Volume 3, Issue 7 (2024).
doi.org/10.1093/pnasnexus/pgae248

\bibitem{Marshall-etal21}
Marshall S. M., Mathis C., Carrick E., Keenan G., 
Cooper G. J. T., Graham H., Craven M., 
Gromski P. S., Moore D. G., Walker S. I., and Cronin L.
Identifying molecules as biosignatures with assembly theory and mass spectrometry, NATURE COMMUNICATIONS | 12:3033 (2021). 
doi.org/10.1038/s41467-021-23258-x

\bibitem{CovTho06}
 Cover T. M. and  Thomas J. A.
Elements of Information Theory,
Wiley-Interscience (2006).


\bibitem{Friederich21}
Friederich S.
Fine-Tuning,
Stanford Encyclopedia of Philosophy (2021)

https://plato.stanford.edu/entries/fine-tuning/

\bibitem{Davies07}
Davies P. C. W. 
The Accidental Universe, Cambridge University Press  (2007).

\bibitem{SmiSmiVer18}
Smith W. S., Smith J. S., and Verducci D., eds. 
Eco-Phenomenology: Life, Human Life, Post-Human Life in the Harmony of the Cosmos, 
Springer,  pp. 131–32 (2018).

\bibitem{Davies03}
Davies P. C. W.  How bio-friendly is the universe, 
Int. J. Astrobiol. 2 (115)
(2003). 
arXiv:astro-ph/0403050. 

\bibitem{BarTip86}
Barrow J. D. and Tipler F. J. 
The Anthropic Cosmological Principle, Oxford University Press (1986).

\bibitem{Carr07}
Carr B. 
Universe or Multiverse? 
Cambridge University Press (2007).

\bibitem{HawHer06}
Hawking S. W. and Hertog, T.  Populating the Landscape: A Top Down Approach, 
Phys. Rev. D73 (12): 123527
(2006).  arXiv:hep-th/0602091v2. 
%Bibcode:2006PhRvD..73l3527H. doi:10.1103/PhysRevD.73.123527. S2CID 9856127.  

\bibitem{Dick15}
Dick S. J. 
The Impact of Discovering Life Beyond Earth,
Cambridge University Press (2015).

\bibitem{Mizrahi17}
Mizrahi M. The Fine-Tuning Argument and the Simulation Hypothesis. Think. 16 (46): 93–102
(2017).  doi:10.1017/S1477175617000094. S2CID 171655427

\bibitem{Adams19}
Adams F. C.  The degree of fine-tuning in our universe – and others, Physics Reports, 807: 1–111. 
(2019). 
doi:10.1016/j.physrep.2019.02.001

%\bibitem{Ball25}
%Ball, P., Why Everything in the Universe Turns More Complex,
%Quanta Magazine, April 2 (2025).




\end{thebibliography}
\end{document}